\documentclass[11pt, a4paper]{scrartcl}
\usepackage[T1]{fontenc}
\usepackage{lmodern}
\usepackage{etoolbox}
\usepackage{amsfonts} 
\usepackage{amsmath}
\usepackage{xcolor} 
\usepackage{hyperref}
\usepackage{placeins}
\usepackage{amssymb} 
\usepackage{multirow}
\usepackage{array}
\newcolumntype{P}[1]{>{\centering\arraybackslash}p{#1}}
\usepackage{makecell}
\usepackage{float}
\usepackage{physics}
\usepackage{braket}
\DeclareUnicodeCharacter{202F}{FIX ME!!!!} 

\usepackage[style=numeric,backend=bibtex,sorting=none,citestyle=numeric-comp]{biblatex}

\bibliography{Refs.bib}

\usepackage{scalerel} 
\usepackage{longtable}
\makeatletter
\patchcmd{\@maketitle}{\LARGE}{\Huge}{\typeout{OK 1}}{\typeout{Failed 1}}
\patchcmd{\@maketitle}{\large \lineskip}{\Large \lineskip}{\typeout{OK 2}}{\typeout{Failed 2}}
\makeatother

\usepackage{datetime}

\newdateformat{monthyeardate}{%
  \monthname[\THEMONTH], \THEYEAR}

\title{
Characterization of Strongly Hyperfine-split Protons by DNP}
\author{Gian-Marco Camenisch, Nino Wili, Gunnar Jeschke, Matthias Ernst}
\date{}

\begin{document}

\maketitle

\section*{Abstract}\label{sec:Abstract}
Dynamic nuclear polarization experiments use microwave irradiation to transfer the larger electron polarization to nuclear spins of interest, and thus enhance the NMR transitions above thermal equilibrium. How the polarization transfer from the electron spin to the nuclear spins in such experiments proceeds and which nuclear spins close to an unpaired electron get polarized and contribute through spin diffusion to the observable bulk nuclear magnetization is not fully understood. We address these questions by combining reverse DNP and band-selective inversion pulses on nuclear spins. We report the electron-detected NMR spectrum of proton spins involved in the direct DNP process in Ox063 trityl samples with protonated and deuterated solvents and variable radical concentrations. We also determine the spin-diffusion barrier surrounding trityl and find that proton spin diffusion is quenched for hyperfine coulings exceding $\sim$ 250 kHz. This corresponds to a radius of the spin diffusion barrier in the range from $5.4$ to $6.8$\,\text{\AA}. Burning a hole into the NMR spectrum of proton spins involved in the direct DNP step reveals an electron-electron spin diffusion process imprinted on the proton spectrum. We explain this diffusion process using a three-spin system consisting of two electron spins and one proton and quantify the electron spin diffusion rate constant.

\section{Introduction}\label{sec:Introduction}
Dynamic nuclear polarization (DNP) is a widely applied method to increase the sensitivity of nuclear magnetic resonance (NMR) spectroscopy by transferring the higher electron polarization to the nuclei of interest. \cite{Abragam1978,LillyThankamony2017,Eills2023} The experiments are typically performed at cryogenic temperatures with a radical embedded in a frozen glassy matrix. \cite{Thurber2010,Lange2012,Takahashi2014,Corzilius2014,Mentink-Vigier2015,Mathies2016} Currently, two different DNP implementations are used in practice: (i) Magic-angle spinning (MAS) DNP operating typically at around 100\;K with the aim to investigate materials, surfaces, and biomolecules. \cite{Becerra1993,Corzilius2012,LillyThankamony2017,Biedenbander2022} (ii) Dissolution DNP typically carried out around 1\;K to generate highly polarized small molecules for metabolic imaging or spectroscopic applications. \cite{Ardenkjær-Larsen2003,Jahnig2016,Jahnig2017,Kurzbach2018,Jannin2019,Elliott2021} Both DNP techniques rely on an efficient spin diffusion process from nuclei (usually protons) near the unpaired electron spin to the bulk nuclear spin bath. \cite{Prisco2021,Stern2021,Stern2023} The nearby nuclei are not directly observable via conventional NMR due to fast relaxation and/or the strong hyperfine coupling to the electron spin. Therefore, it is not well understood how the polarization is transferred from the nearby nuclei to the bulk and where these nuclei are located spatially and spectrally. \cite{Pang2024} Recent studies have demonstrated that the spin diffusion barrier lies within less than 0.6 nm for a trityl radical in a glassy water/glycerol mixture. \cite{Tan2019} Spins within this barrier do not participate in spin diffusion to the bulk due to the large hyperfine splitting. \cite{Smith2012} The spectrum and spin diffusion of proton spins nearby an unpaired electron to bulk protons in TEMPOL embedded in a frozen glycerol-water mixture was demonstrated at 1.4 K using a dissolution DNP setup and conventional NMR bulk spin detection using a modified version of the CEST experiment. \cite{vanZijl2011, Pang2024} Another example is the Nuclear-Nuclear Double Resonance (NUDOR) experiment, which detects strongly coupled protons in different nitroxide biradicals using the cross effect under MAS at 14 T at around 100 K. \cite{Chatterjee2025} Although those are impressive results, the direct NMR detection of nearby spins requires a lot of signal averaging and long measurement times.
In this work, we demonstrate the measurement of an electron-detected spectrum of protons near a trityl radical Ox063 dissolved in protonated and deuterated solvents at 80 K. Electron detection has much higher sensitivity and is thus much faster than a direct detection of the proton spectrum. In particular, the Ox063 trityl radical in a gly-d$_8$:D$_2$O:H$_2$O (6:3:1 by volume) gives high enhancement factors in conventional DNP experiments as first described in Ref. \cite{Mathies2016} and knowledge about this system is of high value. We also report a spectral and spatial diffusion barrier with spin diffusion of protons within the barrier to the bulk protons being quenched. Furthermore, we were able to quantify electron spin diffusion rate constants by burning a hole into the electron-detected proton spectrum and observing its time evolution.

\section{Results and Discussion}\label{sec:Results}
\begin{figure}[H]
\centering
\includegraphics[width=\textwidth]{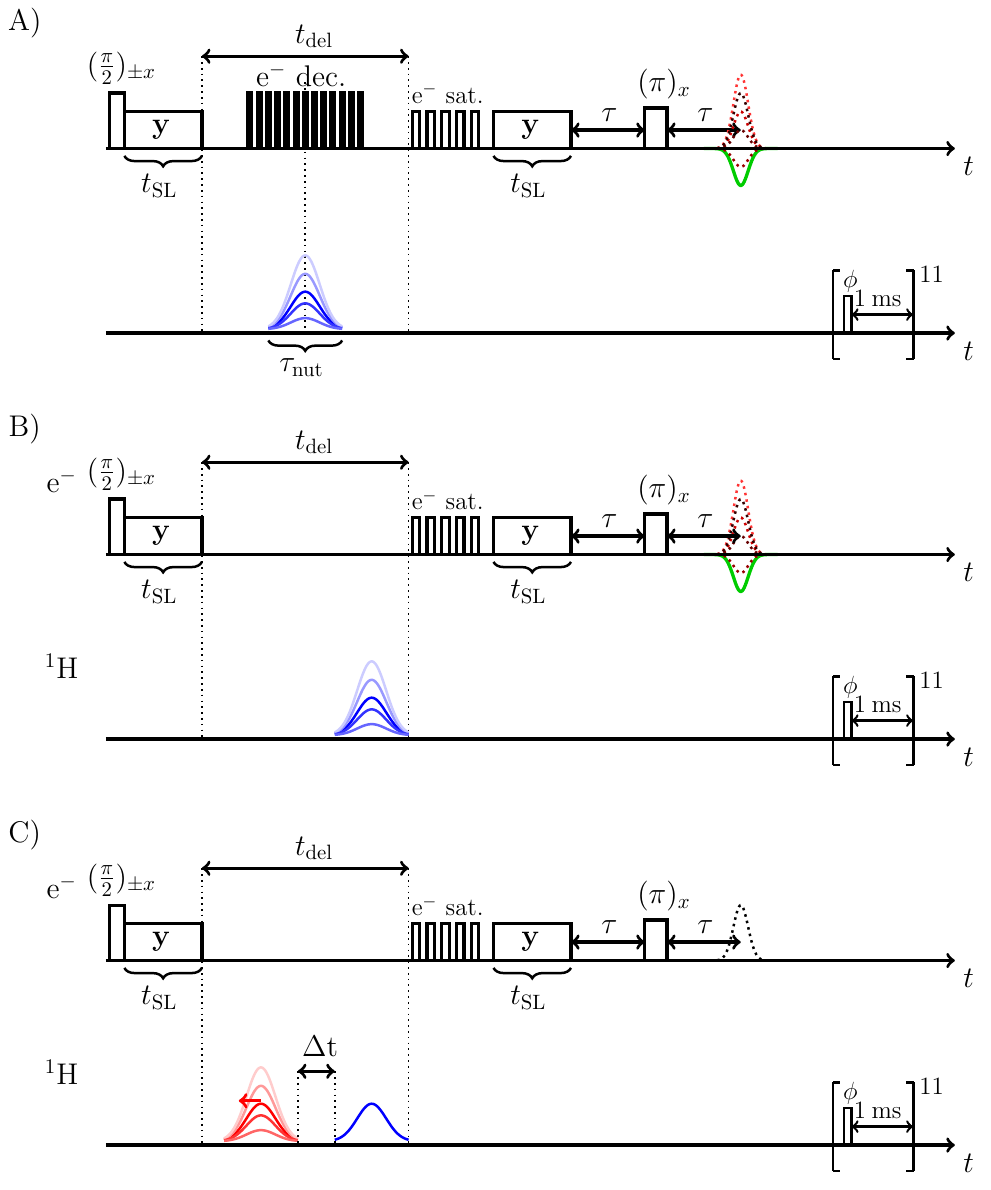}
\caption{Pulse sequences used in the realm of this work. \textbf{A)} is the pulse sequence to measure the electron-detected proton spectra, \textbf{B)} the pulse sequence to monitor the spin diffusion towards the bulk and \textbf{C)} to burn a hole into the proton spectrum. For all experiments a saturation train on the proton spins at the end of each sequence was applied with a carrier frequency equal to the blue nutation pulses and $\phi=100^{\circ}$.}
\label{fig:RevDNPPulseSeq}
\end{figure}

All pulse schemes (Fig. \ref{fig:RevDNPPulseSeq}) used in this work are based on the concept of a reverse DNP contact as first described in Ref. \cite{Wili2022a}. We extend the concept  by introducing pulses on the proton channel.
For both DNP steps the NOVEL sequence \cite{Henstra1988} is used. One of the advantages of using NOVEL as a DNP contact compared to, e.g., the solid effect (SE) \cite{C.D.Jeffries1957} or pulsed DNP sequences \cite{Tan2019b,Redrouthu2022,Wili2022,Redrouthu2023} is that the resonance condition in the NOVEL sequence is satisfied at the center of the EPR line ($\Omega_{\mathrm{0,S}}=0$\,MHz) and thus at the same spectral position as the two pulse Hahn echo detection. This reduces the risk of potential artifacts due to spectral diffusion. Moreover, NOVEL is one of the best performing DNP sequences with respect to the enhancement and available mw power, especially below 3\,T. The NOVEL matching condition $|\nu_{\mathrm{1,S}}|=|\nu_{\mathrm{0,I}}|$ was determined by performing a single NOVEL DNP contact followed by a spin echo as described in Ref. \cite{Wili2022a} and outlined in more detail Sections E.6. and E.7. of the ESI\dag. All pulse sequences have in common that after the initial NOVEL DNP contact and a variable delay $t_{\mathrm{del}}$ the electron spins are saturated to facilitate the back transfer from the polarized proton spins to the electron spin. After the reverse DNP step, the detection of the electron spin is done by a two-pulse Hahn echo sequence. \cite{Hahn1950}

\subsection{Electron-detected proton spectrum}\label{subsec:HSpectrum}
\begin{figure}[H] 
\centering
\includegraphics[width=\textwidth]{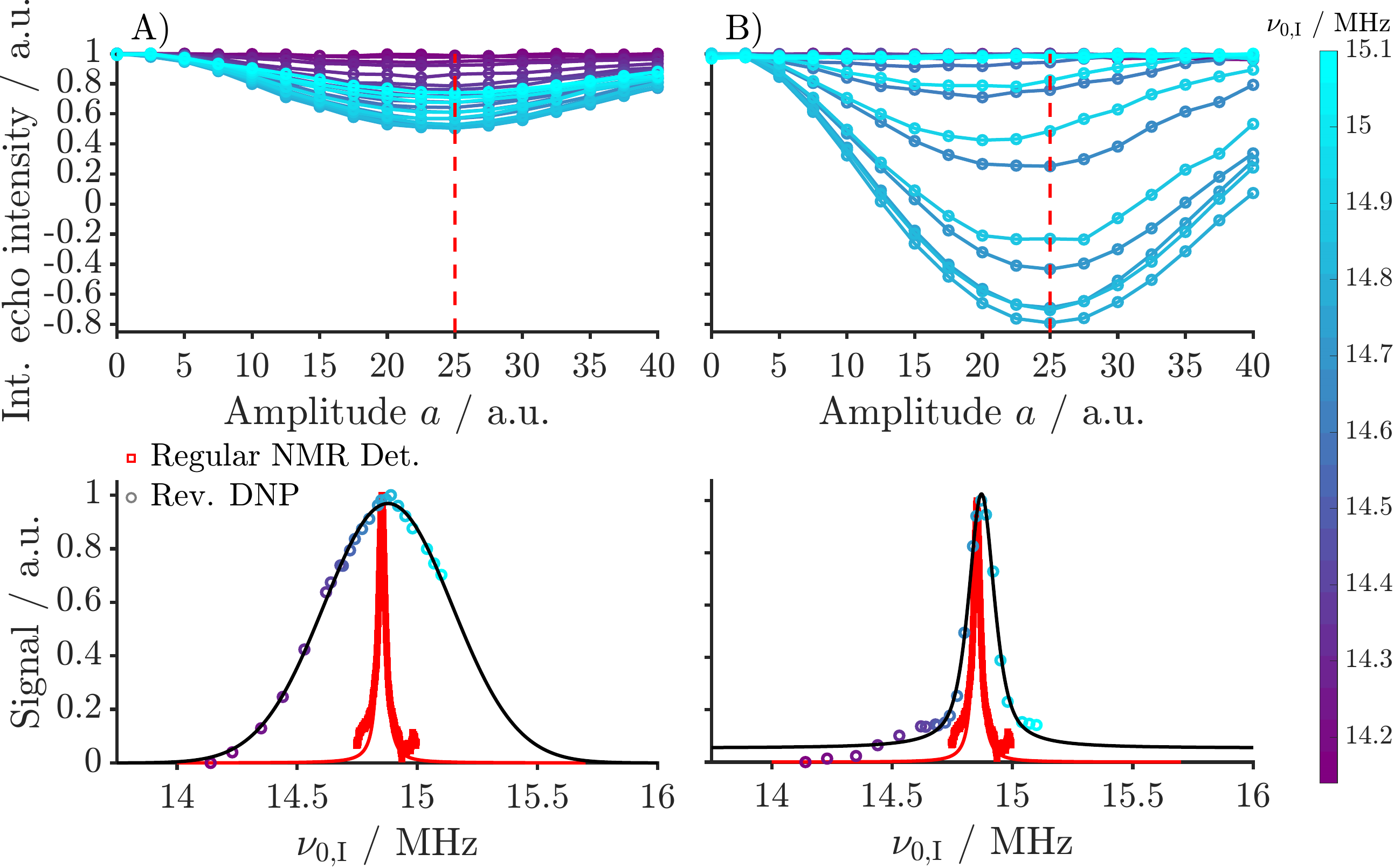}
\caption{Electron-detected proton spectrum for 5\,mM trityl in gly-d$_8$:D$_2$O:H$_2$O (6:3:1) using the pulse sequence of Fig. \ref{fig:RevDNPPulseSeq} A) with $t_{\mathrm{del}}$ = 15\,$\mu$s and $t_{\mathrm{SL}}$ = 4000 ns. The variable-amplitude Gauss pulse was placed in the middle of the delay $t_{\mathrm{del}}$. \textbf{A)} Amplitude nutation traces upon application of a band-selective Gauss pulse of 10\,$\mu$s length. The relative amplitude $a$ of this Gauss pulse is varied from 0 to 40. The first minimum of the amplitude nutation trace is at a relative amplitude of $25$ corresponding to a Rabi frequency of $\sim$ 100 kHz. The carrier frequency of the amplitude nutation trace was varied from 14.14 to 15.10\,MHz throughout different experiments. The minima of the amplitude nutation traces were extracted and plotted against the frequency to obtain the electron-detected proton spectrum. Comparison with the thermal equilibrium signal of the bulk protons using regular NMR detection is shown in red. \textbf{B)} Amplitude nutation traces upon application of the same band-selective Gauss pulse as in A) with simultaneous electron decoupling during that pulse.}
\label{res:fig:CompNutSpectrum_N}
\end{figure}

To record the electron-detected proton spectrum, the pulse sequence shown in Fig. \ref{fig:RevDNPPulseSeq} A) is used. The band-selective variable-amplitude Gauss pulse on the proton channel was placed between the initial DNP contact and the reverse DNP block, i.e., in the middle of the delay $t_{\mathrm{del}}$. The relative amplitude of the Gauss pulse with a pulse length of 10\,$\mu$s was changed from 0 to 40, where a maximum output power of 300 W corresponds to a relative amplitude of 100. The amplitude nutation traces were recorded for carrier frequencies $\nu_{\mathrm{0,I}}$ ranging from 14.14\,MHz to 15.10\,MHz. The Gauss function used for the Gauss pulse has a standard deviation $\sigma=1.96$ $\mu$s. In Fig. \ref{res:fig:CompNutSpectrum_N} we show the result for 5\,mM trityl in gly-d$_8$:D$_2$O:H$_2$O (6:3:1) for $t_{\mathrm{del}}$ = 15\,$\mu$s without electron decoupling A) and with electron decoupling B) for different amplitudes of the variable-amplitude Gauss pulse. We see from the amplitude nutation traces for different carrier frequencies $\nu_{\mathrm{0,I}}$ that the first minimum is found at an amplitude $a=25$ corresponding to $\sim100$ kHz rf-field amplitude which agrees with the theoretical pulse profile as shown in Fig. S1 in the ESI\dag. We clearly see in Fig. \ref{res:fig:CompNutSpectrum_N} A) that we are not able to invert the signal nor do we achieve a negative electron echo. The reason for this is that the proton spectrum is rather broad with a FWHM of $\sim$ 650\,kHz. This effect is also seen in numerical simulations as shown in Fig. S2 in the ESI\dag.\\

The inversion efficiency increases by narrowing the proton spectrum through electron decoupling during the nutation pulse. With electron decoupling an almost complete inversion of the Hahn echo reaching a minimum signal intensity of $\sim-0.8$ in the center of the proton spectrum at $\nu_{\mathrm{0,I}}=14.89$\,MHz. The center frequency is shifted by $\sim$ 40 kHz compared to the thermal equilibrium proton signal from the bulk. The electron-detected proton spectrum is obtained by extracting the first minima of the amplitude nutation traces. This is shown in Fig. \ref{res:fig:CompNutSpectrum_N} A) and B) on the bottom. Without electron-decoupling during the nutation pulse, we obtain an electron-detected proton spectrum with a full-width-half-maximum (FWHM) of $\sim$ 650\,kHz. Note that due to the limited tuning range of the proton circuit, only part of the undecoupled line could be detected. The proton line is by a factor of $\sim20$ wider than for the thermal equilibrium proton signal from the bulk (see Fig. S105 in the ESI\dag). Under electron decoupling during the nutation pulse, the line width of the proton spectrum is reduced by a factor of $\sim5$ compared to the line width without electron decoupling, resulting in a FWHM of $\sim$ 130 kHz. This is still a factor of $\sim4$ larger than the line width of the thermal equilibrium proton signal from the bulk. However, the FWHM of the electron-detected proton spectrum recorded with simultaneous electron decoupling during the pulse is equal to the FWHM of the theoretical inversion bandwidth of the variable-amplitude Gauss pulse, as shown in Fig. S1 in the ESI\dag. The larger linewidth with electron decoupling as compared to direct NMR observation of bulk protons can thus be attributed to power broadening.\\

This motivated us to perform numerical simulations for a more quantitative understanding. For this, we assumed proton lines with FWHM that are equal or smaller than the FWHM of the pulse inversion band. If the proton line width is on the same order as the inversion pulse or even less, a slightly broader proton spectrum is obtained by extracting the minima of the amplitude nutation traces compared to the "true" proton spectrum (see Fig. S3 for a proton FWHM of 30\,kHz and S4 for a proton FWHM of 130 kHz in the ESI\dag). In both cases, the FWHM of the proton spectrum obtained by the minima of the amplitude nutation traces has a FWHM of $\sim$ 130 kHz. That such an approach with extracting the minima of the amplitude nutation traces is suitable to record a proton spectrum that is significantly broader than the bandwidth of the inversion pulse ($\sim$ 130 kHz) is shown in Fig. S5 in the ESI\dag. This shows that our method of indirect measurement of the proton line width works best for spectra that are significantly broader than the theoretical pulse inversion profile. Different types of band-selective rf pulses were tested (I-BURP \cite{Geen1991,Freeman1998}, optimal control designed pulses \cite{Kobzar2005,Gershenzon2007}). Those band-selective pulses have a narrower inversion profile than the 10\,$\mu$s Gauss pulse, but are much longer (up to 100\,$\mu$s or even more). We found that the long duration of such pulses induces substantial loss of signal.\\

Another reason for the relatively broad proton line observed under electron decoupling could be that the electron decoupling is not perfect as can be seen from the shoulder on the left hand side of the main peak in Fig. \ref{res:fig:CompNutSpectrum_N} B) and that the Hahn echo is only inverted up to 80\%. The electron-decoupling consisted of 99 $\pi$-pulses of 12\,ns length spaced by 80.2\,ns with an FWHM of $\sim66.4$\,MHz of the theoretical inversion profile (see Fig. S9 in the ESI\dag). The EPR spectrum of trityl has an FWHM of $\sim6.23$\,MHz. Hence, our pulses should in principle be sufficient to invert the entire EPR spectrum uniformly. Further improvements in the decoupling efficiency can be envisioned by using broadband chirp pulses \cite{Wili2022a} or broadband decoupling sequences as used in solution-state NMR.\cite{Shaka2011}\\

\begin{figure}[H] 
\centering
\includegraphics[width=\textwidth]{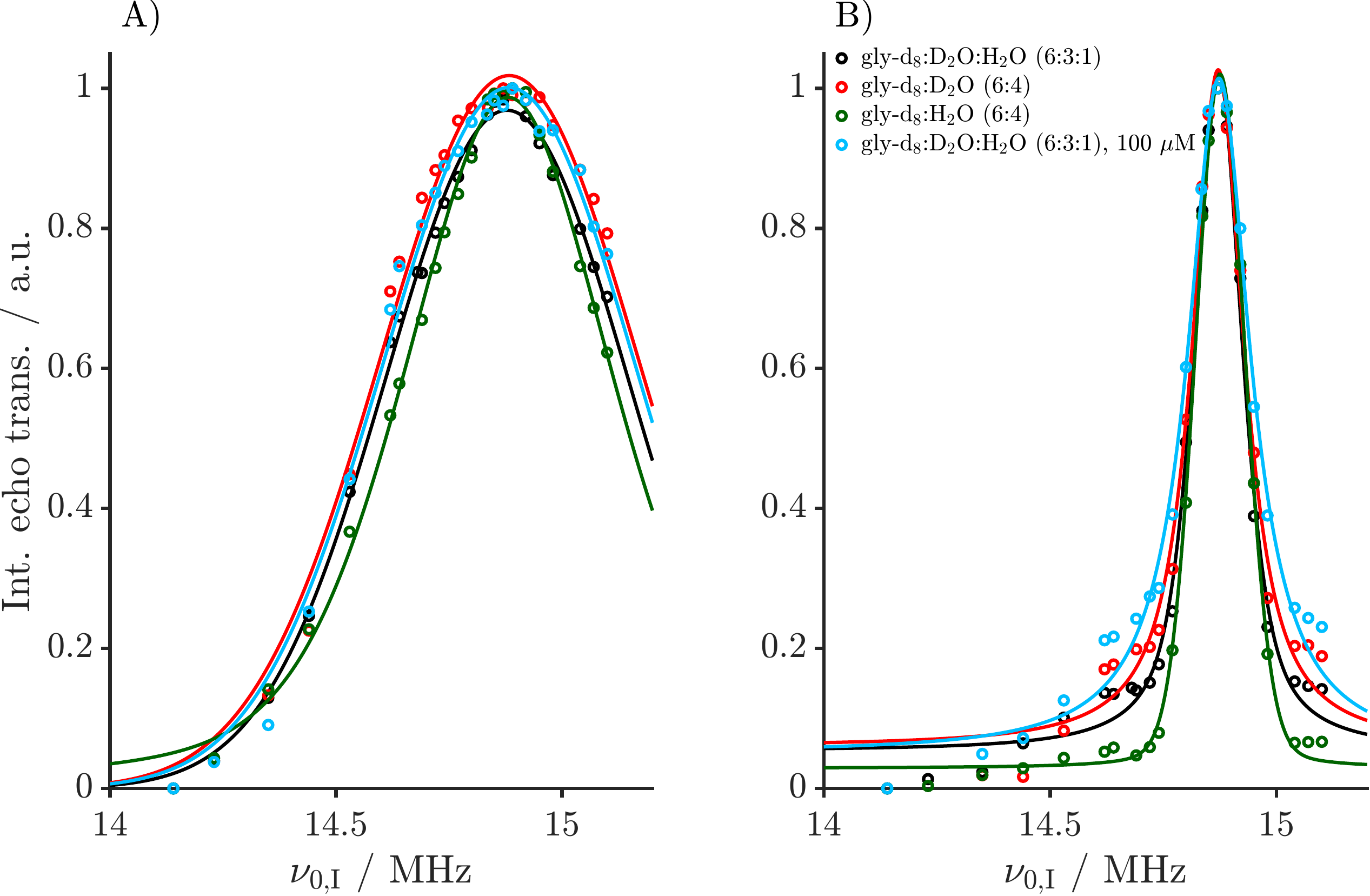}
\caption{Electron-detected proton spectrum for 5\,mM trityl Ox063 in gly-d$_8$:D$_2$O:H$_2$O (6:3:1) black, gly-d$_8$:D$_2$O (6:4) red, gly-d$_8$:H$_2$O (6:4) green and for 100\,$\mu$M trityl Ox063 in gly-d$_8$:D$_2$O:H$_2$O (6:3:1) light-blue using the pulse sequence of Fig. \ref{fig:RevDNPPulseSeq} A). The variable-amplitude Gauss pulse was placed in the middle of the delay $t_{\mathrm{del}}$ = 15\,$\mu$s and $t_{\mathrm{SL}}$ = 4000 ns. Panels \textbf{A)} and \textbf{B)} show the electron-detected proton spectra without and with electron decoupling, respectively. Within experimental error between the four different measurement sessions the electron-detected proton spectra are identical. Please note that the data in black were already shown in Fig. \ref{res:fig:CompNutSpectrum_N}.}
\label{res:fig:CompNutSpectrum_all_15mus}
\end{figure}

In Fig. \ref{res:fig:CompNutSpectrum_all_15mus} we compare the electron-detected proton spectra for samples with different protonation levels of the water, i.e., 5\,mM trityl in gly-d$_8$:D$_2$O:H$_2$O (6:3:1), gly-d$_8$:D$_2$O (6:4) and gly-d$_8$:H$_2$O (6:4) and of different trityl concentration (100\,$\mu$M and 5\,mM both in gly-d$_8$:D$_2$O:H$_2$O (6:3:1)).  In Fig. \ref{res:fig:CompNutSpectrum_all_15mus} \textbf{A)} we show the electron-detected proton spectra without electron decoupling and in Fig. \ref{res:fig:CompNutSpectrum_all_15mus} \textbf{B)} with electron decoupling for the four different sample compositions for a value of $t_{\mathrm{del}}=15$\,$\mu$s. We can see that except for deviations within experimental uncertainty between the four samples, the electron-detected proton spectra are almost identical. For $t_{\mathrm{del}}=800$\,$\mu$s as shown in Fig. S10 in the ESI\dag\; the agreement within the four different data sets is even better. Here, we discuss the experiments for a short time $t_{\mathrm{del}}=15$\,$\mu$s to ensure that spin diffusion to the bulk is negligible (see Chapter \ref{sec:SpinDiffusionBulk}). The observation that the electron-detected proton spectra of the four different samples are almost identical in shape and intensity lets us conclude that we detect protons that are located on the trityl molecule or on close-by solvent molecules rather than the bulk protons. The bulk proton concentration in the four samples is quite different, and we compared the intensity of the thermal equilibrium proton spectra with the theoretical number of protons in the samples (see Tab. S44 in the ESI\dag). The integrated intensities and the number of protons per sample are in good agreement. The calculation for the number of protons in the samples includes the purity of the deuterated molecules and the protons of the trityl molecule. This indicates that there was no contamination during preparation of the samples. Except for the unavoidable proton impurities in gly-d$_8$ ($\sim2$\% protons) and D$_2$O ($\sim0.15$\% protons) as used for the sample preparation of the gly-d$_8$:D$_2$O (6:4) mixture, the amount of protons in the bulk of the matrix of gly-d$_8$:H$_2$O (6:4) as compared to gly-d$_8$:D$_2$O (6:4) is by a factor $\frac{100}{2.15}\sim47$ larger. For such a large difference of numbers in the bulk protons, we would expect a large difference in the electron-detected proton spectra if our method would detect a significant amount of bulk protons (see also thermal equilibrium spectra shown in Fig. S105 and S106 in the ESI\dag).

\subsection{Detection of proton spin diffusion towards the bulk}\label{sec:SpinDiffusionBulk}
\begin{figure}[H] 
\centering
\includegraphics[width=\textwidth]{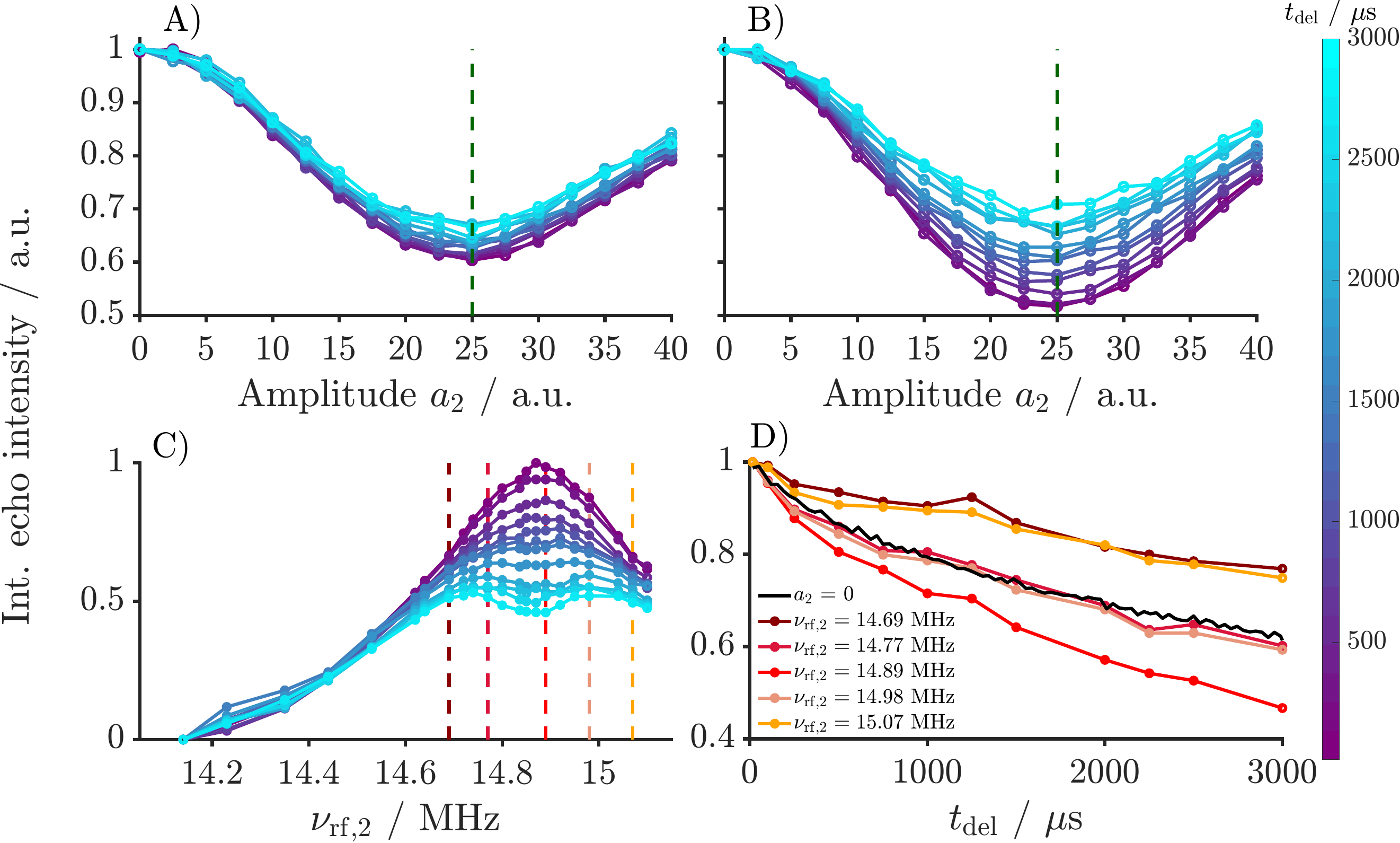}
\caption{Measurement of spin diffusion from protons nearby an electron spin to bulk protons using the pulse sequence of Fig. \ref{fig:RevDNPPulseSeq} B). \textbf{A)} and \textbf{B)} Recorded amplitude nutation traces for $t_{\mathrm{del}}$ ranging from 15\,$\mu$s to 3000\,$\mu$s by varying the amplitude $a_2$ of the Gauss pulse. The sample was 5\,mM trityl in gly-d$_8$:H$_2$O (6:4) and $t_{\mathrm{SL}}$ = 4000 ns. In \textbf{A)} the carrier frequency of the Gauss pulse was $\nu_{\mathrm{rf,2}}=14.77$\,MHz and in \textbf{B)} 14.89\,MHz, with the latter corresponding to the center of the proton spectrum. The data points to plot the proton spectra were extracted along the green dashed line. \textbf{C)} Proton spectra for different delays $t_{\mathrm{del}}$. The spin diffusion towards the bulk is more pronounced in the center of the spectrum. The reddish dashed lines indicate the spectral positions where the data points for the traces in \textbf{D)} were extracted. \textbf{D)} Diffusion traces extracted from \textbf{C)} for different frequencies $\nu_{\mathrm{rf,2}}$. The diffusion to the bulk is fastest for $\nu_{\mathrm{rf,2}}=14.89$\,MHz corresponding to the center of the spectrum and gets slower moving away from the center. The black solid line indicates a measurement in the absence of any Gauss pulse i.e. $a_2 = 0$.}
\label{res:fig:SubPlot_Bulk_Diffusion_4000ns_SL}
\end{figure}

In a second step, we investigated the decay of proton polarization during the delay $t_{\mathrm{del}}$ (see Fig. \ref{fig:RevDNPPulseSeq} B)). This kind of experiment was already reported in Ref. \cite{Wili2022a} for a different matrix composition and different trityl concentrations at Q-band (1.2 T field) without pulses on the proton channel. However, our goal here is to measure the decay of proton polarization  as a function of the proton frequency $\nu_{\mathrm{rf,2}}$. Therefore, the Gauss pulse to record the frequency-selective proton spectrum was placed at the end of the delay $t_{\mathrm{del}}$ as shown in Fig. \ref{fig:RevDNPPulseSeq} \textbf{B)}. The delay $t_{\mathrm{del}}$ was increased from 15\,$\mu$s to 3000\,$\mu$s using a NOVEL contact time of $t_{\mathrm{SL}}=$ 4000 ns for both DNP contacts. The amplitude nutation traces were recorded for a range of frequencies from $\nu_{\mathrm{rf,2}}=$ 14.14 to 15.1\,MHz for 5\,mM trityl in gly-d$_8$:H$_2$O (6:4). Figure \ref{res:fig:SubPlot_Bulk_Diffusion_4000ns_SL} \textbf{A)} shows the measurement for $\nu_{\mathrm{rf,2}}=14.77$\,MHz and \textbf{B)} for $\nu_{\mathrm{rf,2}}=14.89$\,MHz. The reconstructed proton spectra for different delays $t_{\mathrm{del}}$ are shown in Fig. \ref{res:fig:SubPlot_Bulk_Diffusion_4000ns_SL} \textbf{C)}. They were obtained by extracting the minima of the amplitude nutation traces at an amplitude $a_2=25$, which is indicated by the green dashed lines in Fig. \ref{res:fig:SubPlot_Bulk_Diffusion_4000ns_SL} \textbf{A)} and \textbf{B)}. One can clearly see that the proton polarization decays and that this decay is more pronounced in the center of the spectra, i.e., in the range of $14.64\;\mathrm{MHz}\leq \nu_{\mathrm{rf,2}}\leq15.07$\,MHz. Outside this range no significant decay of the proton polarization can be observed on this time scale. Comparing this observation with the thermal bulk proton spectrum as shown in Fig. S105 in the ESI\dag\; with a center frequency $\sim \nu_{\mathrm{0,I}} = 14.853$\,MHz and a FWHM = 0.054\,MHz, we attribute this loss of magnetization to spin diffusion to the bulk protons rather than a decay originating from longitudinal relaxation of the protons (bulk proton relaxation times $T_{\mathrm{1,H}}$ are on the order of seconds for all samples). Outside this range spin diffusion to the bulk protons is much slower due to the large difference in the proton frequency between the bulk and the protons nearby the electron spin. For a shift in frequency of $14.853\,\mathrm{MHz} - 14.6\,\mathrm{MHz}=0.253\,\mathrm{MHz}$ the electron-proton distance can be calculated to define a spatial diffusion barrier. For an angle $\theta$ between the external magnetic field vector $\Vec{B}$ and the electron-proton distance vector $\vec{r}_{\mathrm{SI}}$ we find for $\theta = 0^{\circ}$ $|\Vec{r}_{\mathrm{SI}}| = 6.8 $\,\text{\AA} and for $\theta = 90^{\circ}$ $|\Vec{r}_{\mathrm{SI}}| = 5.4 $\,\text{\AA}. This is in very good agreement with the diffusion barrier of $<6$\,\text{\AA} as reported by Tan et al. \cite{Tan2019}.
The spin diffusion to the bulk protons is faster at the center of the proton line ($\nu_{\mathrm{rf,2}}=$ 14.89\,MHz) than at the edge of the active range ($\nu_{\mathrm{rf,2}}=$ 14.64\,MHz or $\nu_{\mathrm{rf,2}}=$ 15.07\,MHz) as can be seen in Fig. \ref{res:fig:SubPlot_Bulk_Diffusion_4000ns_SL} \textbf{D)}. In the absence of any nutation pulses ($a_2 = 0$), an average decay is measured due to the missing proton frequency selectivity.

A simple fit to a mono-exponential or stretched exponential decay as shown in Fig. \ref{res:fig:SubPlot_Bulk_Diffusion_4000ns_SL} \textbf{D)} did not give satisfactory or stable results. A longer time scale would be required to unambiguously characterize the decay constant of the spin diffusion process. In reality a sum of mono-exponential or stretched exponential functions might describe the spin diffusion accurately, similar to Ref. \cite{Stern2021}. Also the electron spin polarization plays a role in the nuclear spin diffusion. \cite{Stern2023} It was found experimentally and explained using Lindblads Master equation in Ref. \cite{Stern2023} that the lower the electron spin polarization the faster the nuclear spin diffusion. A detailed investigation of the spin diffusion for different matrix compositions and electron concentrations is beyond the scope of this article, but would be an interesting research topic. In the context of this work the aim is to quantify the spectral range in the proton spectrum where spin diffusion to the bulk takes place. In Fig. S11 in the ESI\dag\; we show the experimental data for a NOVEL contact time $t_{\mathrm{SL}}$ of 800\,ns. The protons nearby the electron spin are thus less polarized, and the observed spin diffusion towards the bulk is slower. This indicates that a significant amount of the observed loss in polarization for increasing $t_{\mathrm{del}}$ can be attributed to spin diffusion towards the bulk rather than to spin relaxation of the nuclei. The latter is expected to be independent of the polarization level of the nearby protons. With our experimental data as shown in Fig. \ref{res:fig:SubPlot_Bulk_Diffusion_4000ns_SL} we observe a spectral region ranging from $\sim$ 14.64\,MHz to 15.07\,MHz in which spin diffusion to bulk protons takes place. Outside this spectral region spin diffusion to the bulk is quenched due to the strong HFI.

\subsection{Electron-spin diffusion imprinted on the proton spectrum}
 In a final experiment (Fig. \ref{fig:RevDNPPulseSeq} \textbf{C)}) the pulse sequence has an additional band-selective Gauss pulse to burn a hole into the electron-detected proton spectrum. The first pulse drawn in red is the hole burning pulse with a variable amplitude $a_1$ to ensure optimal hole-burning efficiency at a fixed carrier frequency $\nu_{\mathrm{rf,1}}$. The blue pulse is the detection pulse to record the proton spectrum and has an amplitude $a_2$ and a carrier frequency $\nu_{\mathrm{rf,2}}$. To maintain optimal detection excitation and sensitivity, the NMR coil of the resonator is tuned and matched with respect to the carrier frequency of the detection pulse. The amplitude $a_2$ of this pulse is kept constant during the experiment. The carrier frequency $\nu_{\mathrm{rf,2}}$ of the detection pulse is systematically varied across the proton spectrum from 14.14\,MHz to 15.10\,MHz. Tuning and matching was always on resonance with respect to the detection pulse to maintain the same detection sensitivity across the proton spectrum. Here we discuss the experiments where the hole was burned at $\nu_{\mathrm{rf,1}} = 14.69$\,MHz and $\nu_{\mathrm{rf,1}} = 14.89$\,MHz. Experiments with a hole burned at $\nu_{\mathrm{rf,1}} = 15.04$\,MHz can be found in the ESI\dag, Fig. S73. At large differences between $\nu_{\mathrm{rf,1}}$ and $\nu_{\mathrm{rf,2}}$, the amplitude of the hole burning pulse would no longer be the same as the one of the second pulse. To account for that, the amplitude $a_1$ of the hole-burning pulse is incremented during the experiments. In Fig. S13 in the ESI\dag\; we show such an experimental amplitude trace for variation of $a_1$ for different frequencies $\nu_{\mathrm{rf,2}}$ and $\Delta t$ for the sample 5\,mM trityl in gly-d$_8$:D$_2$O:H$_2$O (6:3:1) and $\nu_{\mathrm{rf,1}}=14.69$\,MHz. The amplitude traces of all other samples and frequencies $\nu_{\mathrm{rf,1}}$ can be found in Sections C.1. - C.5. of the ESI\dag. To observe time-dependent changes in the spectra, the delay between the two Gauss pulses $\Delta t$ was incremented independently. All details about the data analysis and processing of the experimental data can be found in Section \ref{sec:Materials}.
\\
 
\begin{figure}[H] 
\centering
\includegraphics[width=\textwidth]{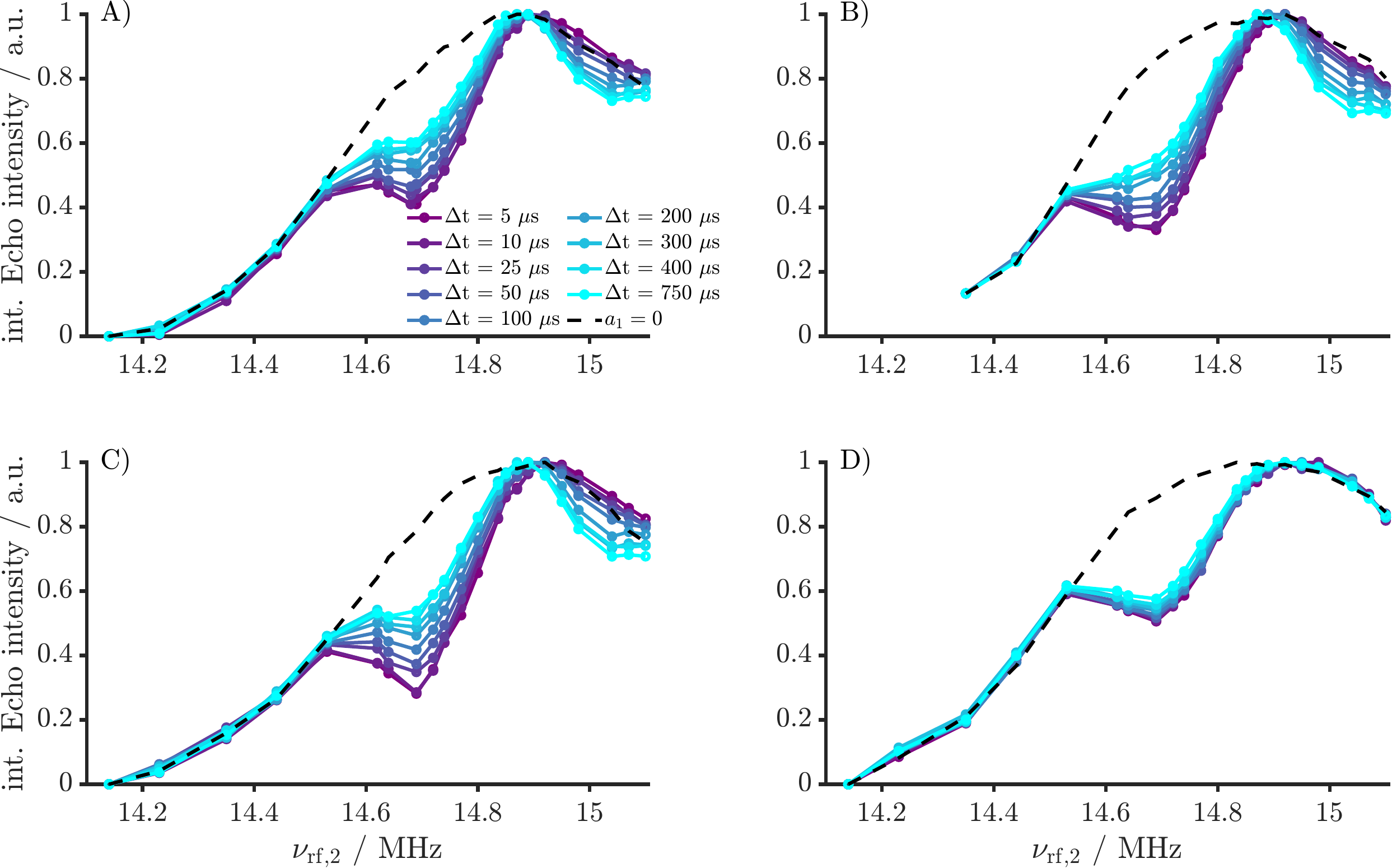}
\caption{Hole burning spectra for delays $\Delta t$ between the two Gauss pulses varying from 5 to 750\,$\mu$s, $t_\mathrm{del}=$ 800\,$\mu$s and $t_\mathrm{SL}=$ 4000 ns. The hole was burned at $\nu_{\mathrm{rf,1}}=14.69$\,MHz. The black dashed lines are obtained by using the data points with a$_1$ = 0 i.e. in the absence of any hole burning pulse. The sample in subplot A) is 5\,mM trityl in gly-d$_8$:D$_2$O:H$_2$O (6:3:1), in subplot B) 5\,mM trityl in gly-d$_8$:D$_2$O (6:4), in subplot C) 5\,mM trityl in gly-d$_8$:H$_2$O (6:4) and in subplot D) 100\,$\mu$M trityl in gly-d$_8$:D$_2$O:H$_2$O (6:3:1).}
\label{res:fig:Diff_14_69_Comp}
\end{figure}

  Figure \ref{res:fig:Diff_14_69_Comp} shows the proton spectra as a function of $\Delta t$ for a hole burning frequency of $\nu_{\mathrm{rf,1}}=14.69$\,MHz for four different samples. In all spectra, it can be observed that the hole burning was successful and that the hole is slowly filled for larger values of $\Delta t$. We observe an intensity loss symmetrically shifted around the center of the proton spectrum at $\nu_{\mathrm{rf,2}}\sim14.89$\,MHz. The hole in the proton spectrum has a width of $\sim$ 320 kHz which matches the width of the theoretical inversion profile of $\sim$ 360 kHz as shown in Fig. S1.
  
\begin{figure}[H] 
\centering
\includegraphics[width=\textwidth]{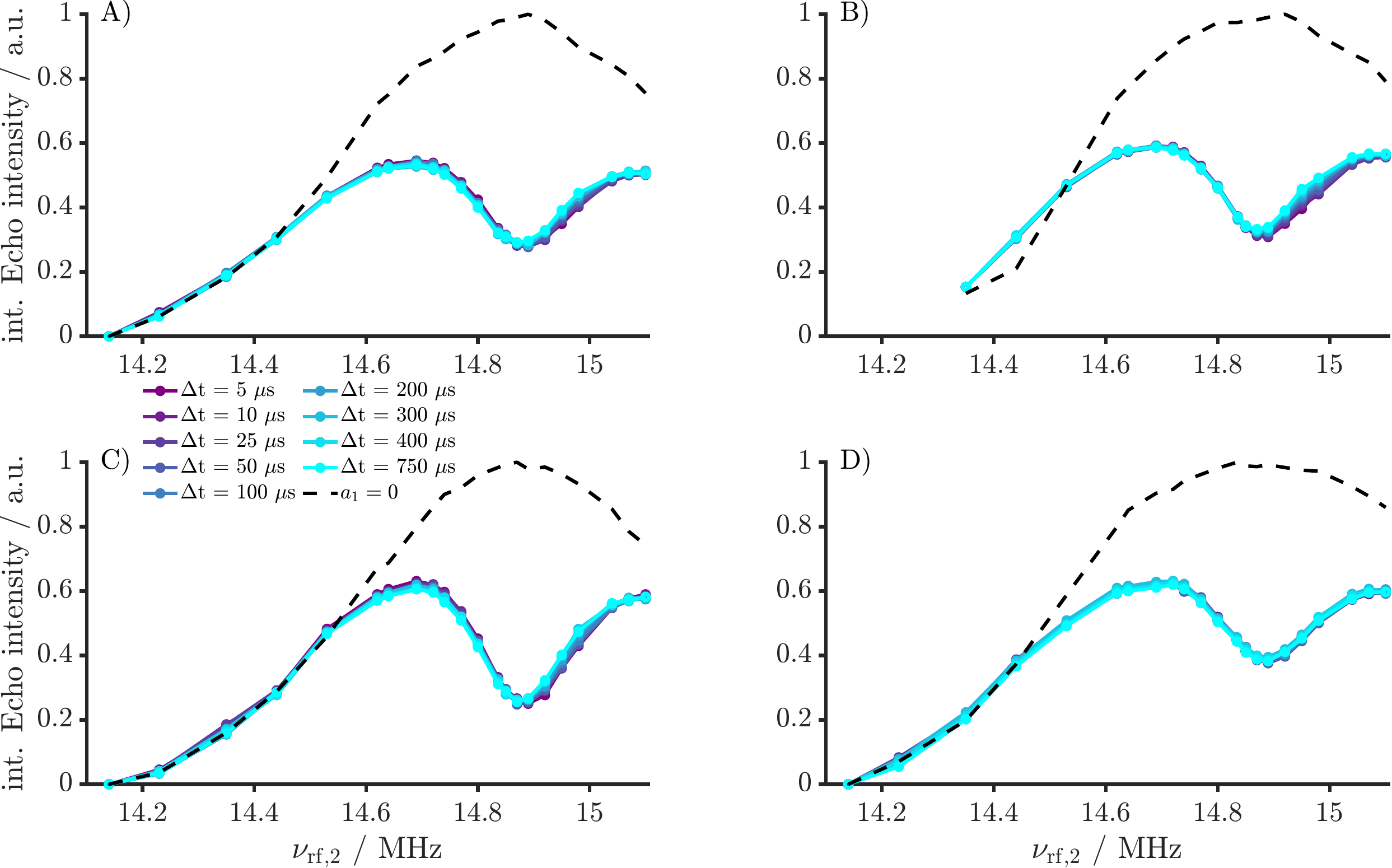}
\caption{Hole burning spectra for delays $\Delta t$ between the two Gauss pulses varying from 5 to 750\,$\mu$s, $t_\mathrm{del}=$ 800\,$\mu$s and $t_\mathrm{SL}=$ 4000 ns. The hole was burned at $\nu_{\mathrm{rf,1}}=14.89$\,MHz. The black dashed lines are obtained by using the data points with $a_1$ = 0 i.e. in the absence of any hole-burning pulse. The sample in subplot A) is 5\,mM trityl in gly-d$_8$:D$_2$O:H$_2$O (6:3:1), in subplot B) 5\,mM trityl in gly-d$_8$:D$_2$O (6:4), in subplot C) 5\,mM trityl in gly-d$_8$:H$_2$O (6:4) and in subplot D) 100\,$\mu$M trityl in gly-d$_8$:D$_2$O:H$_2$O (6:3:1).}
\label{res:fig:Diff_14_89_Comp}
\end{figure}

\begin{figure}[H] 
\centering
\includegraphics[width=\textwidth]{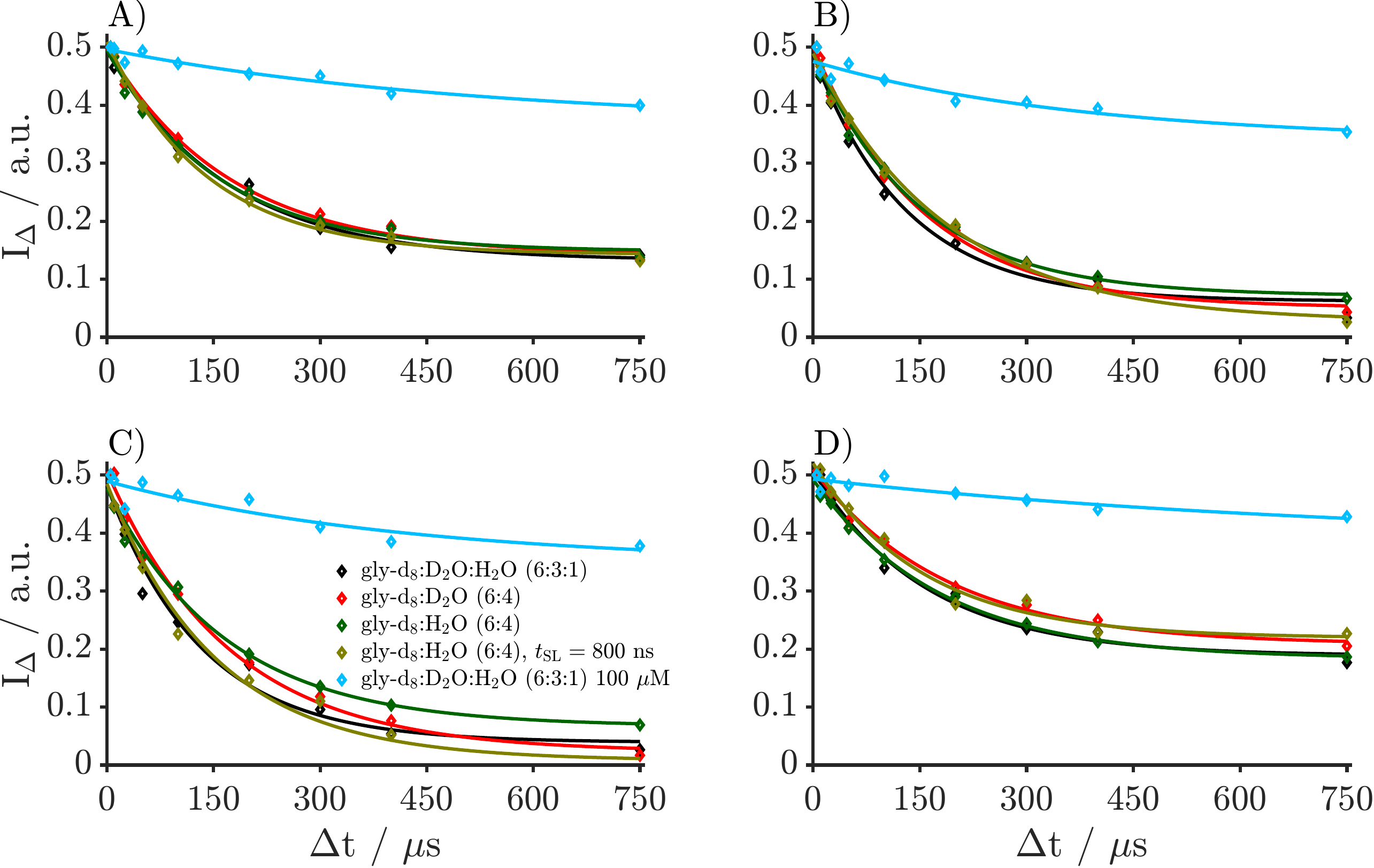}
\caption{Comparison of the difference magnetization according to Eq. (\ref{res:eq:DiffMagnetization}) for different combinations of $\nu_{\mathrm{h}}$ and $\nu_{\mathrm{f}}$ across the four different samples. The hole is burned at $\nu_{\mathrm{rf,1}} = 14.69$\,MHz. A) corresponds to $\nu_{\mathrm{h}}=$14.69\,MHz, $\nu_{\mathrm{f}}=$15.04\,MHz, B) to $\nu_{\mathrm{h}}=$14.77\,MHz, $\nu_{\mathrm{f}}=$14.98\,MHz, C) to $\nu_{\mathrm{h}}=$14.80\,MHz, $\nu_{\mathrm{f}}=$14.95\,MHz and D) to $\nu_{\mathrm{h}}=$14.64\,MHz, $\nu_{\mathrm{f}}=$15.10\,MHz.}
\label{fig:DiffMagn_14_69_all}
\end{figure}

  This diffusion process is observed in samples with 5\,mM trityl concentration with different degrees of protonation as shown in Fig. \ref{res:fig:Diff_14_69_Comp} \textbf{A)} to \textbf{C)}. It is strongly reduced in the sample with a trityl concentration of 100\,$\mu$M. This constitutes first evidence that the observed process includes a spin system consisting of two or more electron spins. The proton polarization transport process occurs on a time scale of $\sim 100$\,$\mu$s (see Fig. \ref{fig:DiffMagn_14_69_all} and Tab. \ref{tab:Comp_Sum_Diff_Magn_14_69_all}), while the relaxation time of the electrons is $T_{\mathrm{1,e}}\sim$ 2.5 ms for all samples regardless of the trityl or proton concentration. The bulk proton relaxation times $T_{\mathrm{1,H}}$ are on the order of seconds for all samples, and even protons near the radicals are unlikely to relax so fast. \cite{Abragam1978} The hole forming at $\nu_{\mathrm{rf,2}}\sim15.04$\,MHz broadens over time, which is another indication that the observed spin transport process is not caused by $T_1$ relaxation. The latter would only reduce the depth of the hole over time, but not influence its width.  The process is symmetric for holes burned at $\nu_{\mathrm{rf,1}}=15.04$\,MHz and $\nu_{\mathrm{rf,1}}=14.69$\,MHz (see Fig. S73 in ESI\dag). For a hole burned at the center of the spectrum at $\sim \nu_{\mathrm{rf,1}}=14.89$\,MHz we do not observe any spectral transport process at all, as shown in Fig. \ref{res:fig:Diff_14_89_Comp}.
  \\
  
  These observations can be explained by a simple three-spin model consisting of two coupled electron spins and one proton spin. The proton spin needs to be hyperfine coupled to one of the electron spins. If the two electrons undergo a flip-flop transition (electron spin diffusion), the hyperfine coupling to the proton changes sign, and the frequency of the proton changes. If the energy difference between the two electron spins is roughly equal to the hyperfine coupling, the process in the three-spin system would be energy conserving and, therefore, fast. This is what we observe experimentally when burning a hole at $\nu_{\mathrm{rf,1}}=14.69$\,MHz or $\nu_{\mathrm{rf,1}}=15.04$\,MHz. The hole initially burned at $\nu_{\mathrm{rf,1}}$ is slowly filled with magnetization originating from the spectral region symmetric around the center of the proton spectrum. In the center of the proton spectrum at $\sim \nu_{\mathrm{rf,1}}=14.89$\,MHz the electron spin diffusion is not visible on the proton spectra because the hyperfine coupling is too small to generate a visible splitting of the resonances. Thus, the exchange of magnetization between the two spectral positions in the proton spectrum is most likely caused by an electron spin-diffusion process. Similar phenomena were observed in MAS NMR in the \textsuperscript{13}C spectra of adamantane \cite{Ernst1998} or in \textsuperscript{13}C spectra of a small organic molecule with trifluoromethyl group.\cite{Bartalucci2025}

To check that magnetization is indeed transported symmetrically across the center of the proton line, one can analyze the behavior of the sum $\mathrm{I}_{\Sigma}(\Delta t)$ and difference magnetization $\mathrm{I}_{\Delta}(\Delta t)$ for two spectral positions $f_{\mathrm{h}}$ and and $f_{\mathrm{f}}$. The frequency $f_{\mathrm{h}}$ denotes the spectral position near the hole and $f_{\mathrm{f}}$ the symmetric position where magnetization is lost. The two frequencies are symmetric around the center of the spectrum, i.e., $f_{\mathrm{h}} = \nu_{\mathrm{0,I}}+\Delta \nu$ and $f_{\mathrm{f}} = \nu_{\mathrm{0,I}}-\Delta \nu$. Then the sum magnetization $\mathrm{I}_{\Sigma}$ and the difference magnetization $\mathrm{I}_{\Delta}$ can be defined as 
\begin{align}
    \mathrm{I}_{\Sigma}(\Delta t) &= \frac{1}{2}\left(\mathrm{I}_{\mathrm{h}}\left(\Delta t\right)+\mathrm{I}_{\mathrm{f}}\left(\Delta t\right)\right)/\left(\mathrm{I}_{\mathrm{h}}\left(0\right)+\mathrm{I}_{\mathrm{f}}\left(0\right)\right)\label{res:eq:SumMagnetization}, \\
    \mathrm{I}_{\Delta}(\Delta t) &= \frac{1}{2}\left(\mathrm{I}_{\mathrm{f}}\left(\Delta t\right)-\mathrm{I}_{\mathrm{h}}\left(\Delta t\right)\right)/\left(\mathrm{I}_{\mathrm{f}}\left(0\right)-\mathrm{I}_{\mathrm{h}}\left(0\right)\right), \label{res:eq:DiffMagnetization}
\end{align}
 where $\mathrm{I}_{\mathrm{h}}\left(\Delta t\right)$ and $\mathrm{I}_{\mathrm{f}}\left(\Delta t\right)$ represent the intensities of the proton spectra at position $f_{\mathrm{h}}$ and $f_{\mathrm{f}}$ as shown in Fig. \ref{res:fig:Diff_14_69_Comp}. If the magnetization is transferred symmetrically around the center of the spectrum, the sum magnetization $\mathrm{I}_{\Sigma}$ is constant over $\Delta t$ and the difference magnetization $\mathrm{I}_{\Delta}$ decays towards a plateau $I_{\infty}$. Please note that the intensity $\mathrm{I}_{\mathrm{h}}\left(\Delta t=0\right)$ was set to the intensity at $\Delta t$ = 5\,$\mu$s. In Sections C.1. - C.5. of the ESI\dag\; we show the comparison of $\mathrm{I}_{\Sigma}$ and $\mathrm{I}_{\Delta}$ for all four different samples and combinations of $f_{\mathrm{h}}$ and and $f_{\mathrm{f}}$. The sum magnetization $\mathrm{I}_{\Sigma}$ is indeed constant over the time scale of $\Delta t$ for almost all cases. An exception is the case for the sample with 100\,$\mu$M trityl in gly-d$_8$:D$_2$O:H$_2$O (6:3:1), where almost no diffusion process is visible. In Fig. \ref{fig:DiffMagn_14_69_all} we only show the difference magnetization for the hole burned at $\nu_{\mathrm{rf,1}} = 14.69$\,MHz for all four different samples. We fitted the difference magnetization $\mathrm{I}_{\Delta}\left(\Delta t\right)$ to an exponential decaying function including a plateau $I_{\infty}$ as described in Eq. (S.3) in the ESI\dag. The extracted fitting parameters are given in Tab. \ref{tab:Comp_Sum_Diff_Magn_14_69_all}. From the data in Fig. \ref{fig:DiffMagn_14_69_all} and Tab. \ref{tab:Comp_Sum_Diff_Magn_14_69_all} we can see that the three samples with 5\,mM trityl concentration show almost identical decays. A small variation in the decay-rate constant and the final value with respect to the proton concentration can be observed especially at the center of the line. Among the samples with 5\,mM trityl concentrations, the deuterated sample shows the slowest decay and the sample with only H$_2$O the fastest decay. Here we used a simple mono-exponential function to fit the data. A sum of exponentials or stretched exponentials might be a better fitting model. We refrain from optimizing fit quality by making the model more complex, as we aim to describe the origin of the diffusion process only qualitatively here. 
 
 When we compare the 5\,mM samples to the 100\,$\mu$M sample, we see a drastic slowdown of the spectral diffusion process by a factor of $4-5$ in the decay rate constant $\tau$ upon reducing electron concentration by a factor of 50. This is consistent with our initial assessment that the observed proton polarization transport is the result of an electron spin diffusion process within a three-spin system consisting of two coupled electron spins and on proton spin. A decrease in the electron-electron coupling strength between the 5\,mM samples and the 100\,$\mu$M sample is also visible in phase memory time $\mathrm{T}_{\mathrm{M}}$ measured using a two pulse Hahn echo and shown in Fig. S90 in the ESI\dag.
 
\begin{table}[H]                                         
\renewcommand{\arraystretch}{1.5}
\centering                                    
\begin{tabular}{|P{6cm}|P{2cm}|P{2cm}|P{2cm}|}
\hline                                          
Sample & $\Delta I$ / a.u. & $\tau$ / $\mu$s & $I_{\infty}$ / a.u. \\            
\hline                                         
gly-d$_8$:D$_2$O:H$_2$O (6:3:1) 5\,mM & 0.361 & 168.499 & 0.132 \\
\hline                                        
gly-d$_8$:D$_2$O (6:4) 5\,mM & 0.356 & 172.961 & 0.142 \\
\hline                                
gly-d$_8$:H$_2$O (6:4) 5\,mM / 4000 ns & 0.347 & 155.591 & 0.148 \\
\hline                                         
gly-d$_8$:H$_2$O (6:4) 5\,mM / 800 ns & 0.367 & 140.626 & 0.142 \\
\hline                                           
gly-d$_8$:D$_2$O:H$_2$O (6:3:1) 100\,$\mu$M & 0.129 & 530.830 & 0.367 \\                    
\hline                                          
\end{tabular}                                      
\caption{Comparison of the fit parameters from a mono-exponential decay as given in Eq. (S.3) in the SI and shown in Fig. \ref{fig:DiffMagn_14_69_all} for difference magnetization $\mathrm{I}_{\Delta}$ for all four samples. The hole was burned at $\nu_{\mathrm{rf,1}} = 14.69$\,MHz and $\nu_{\mathrm{h}}=$14.69\,MHz, $\nu_{\mathrm{f}}=$15.04\,MHz.}
\label{tab:Comp_Sum_Diff_Magn_14_69_all}
\end{table}

\section{Conclusion}\label{sec:Conclusion}

We have demonstrated that it is possible to indirectly detect the protons close to electron spins using the reverse DNP scheme and frequency-band selective proton pulses. Substantial spectral splittings caused by the large hyperfine couplings make direct detection via regular NMR experiments difficult. We show experimentally that the observed proton line originates from protons very close to the electron spins most likely to a large extent from the protons on the trityl molecule itself. This can be inferred from the fact that the intensity of the electron-detected spectrum is largely independent of the proton concentration of the solvent. The electron-detected proton spectrum is also independent of the trityl concentration.\\

The center of the electron-detected proton spectrum is shifted by $\sim$ 40 kHz and the FWHM increases by a factor $\sim$ 20 compared to the bulk proton signal. We validated our experimental results by simultaneous application of electron decoupling during the band-selective inversion pulse, which reduces the proton line width significantly. Under optimized $\pi$-pulse decoupling on the electrons, the FWHM is reduced by a factor of 5. The observed proton line width under electron decoupling is still a factor $\sim4$ larger than the line width of the bulk spectrum. This is most likely due to the large radiofrequency field amplitude of the band-selective proton pulses. In addition, imperfect electron decoupling might also contribute to the larger line width, and we expect that the decoupling efficiency could be increased by using broadband chirp pulses.\\  

By measuring the spin-diffusion rate constant to the bulk as a function of the proton spectral frequency, we were able to characterize the spectral diffusion barrier of the proton spins close to the electrons. Thus, we observe a spectral region ranging from $\sim$ 14.64\,MHz to 15.07\,MHz in which spin diffusion to bulk protons centered at $\nu_{\mathrm{0,I}}\sim$ 14.85\,MHz takes place. This results in a radius for the diffusion barrier ranging from $\sim$ 5.4 to 6.8\,\text{\AA}. For proton spins with a electron-proton distance shorter than $\sim$ 5.4\,\text{\AA}, diffusion to the bulk protons is quenched. A straightforward analysis of the diffusion time constant is complicated by different effects that influence the diffusion of polarization from protons close to an electron spin to the bulk protons.\\

Using hole-burning experiments on the proton line, we were able to characterize electron spin diffusion  through a change in the proton resonance frequency. The electron spin diffusion mechanism was confirmed by comparing samples with two different concentrations of trityl. The electron spin diffusion process is absent in a sample of 100\,$\mu$M trityl in gly-d$_8$:D$_2$O:H$_2$O (6:3:1) and is almost independent of the concentration of protons in the matrix.\\

The observations in this publication have implications for the future design of new DNP experiments and probably also ENDOR experiments. We were able to detect the proton spectrum of protons spins nearby an unpaired electron spin in trityl Ox063 and to reduce the line width of that spectrum by applying electron-decoupling. Furthermore, we showed that a hole can be burned into the spectrum and an electron-electron spin diffusion process within an electron-electron-proton spin system can be observed. More important, we showed that not all protons nearby an electron spin take part in the spin diffusion process towards the bulk proton spin bath. This has important consequences for the design of more efficient DNP radicals.
In this work, we measured a narrow-line trityl radical. We expect that the experiments in this work can be extended by pulsed broadband DNP sequences  \cite{Nielsen2024,Niccoli2026} to also investigate radicals with broader EPR spectra such as nitroxides.

\section{Materials and Methods}\label{sec:Materials}
\subsection{Sample Preparation}\label{subsec:SamplePrep}
The molecular weights and densities used to prepare the four different samples are listed in Tab. \ref{mat:tab:ChemicalCompounds}.
\begin{table}[H]                                              
\renewcommand{\arraystretch}{1.5}                 
\centering                                              
\begin{tabular}{|P{3cm}|P{3cm}|P{3cm}|}                                 
\hline                                              
Compound & MW / (g/mol) & $\rho_0$ / (g/mL) \\                                      
\hline                                             
Ox063 trityl & 1359 & - \\        
\hline
H$_2$O & 18.02 & 0.997 \\        
\hline 
D$_2$O & 20.03 & 1.11 \\        
\hline
glycerol-d$_8$ & 100.14 & 1.371 \\        
\hline
\end{tabular}                                              
\caption{List of molecular weight (MW) and density $\left(\rho_0\right)$ used for sample preparation. Ox063 trityl is a solid and therefore no density is listed.}
\label{mat:tab:ChemicalCompounds}        
\end{table}
The four different samples were prepared by the following procedures:\newline
\textbf{5\,mM trityl in gly-d$_8$:D$_2$O:H$_2$O (6:3:1)}\newline
1.69\,mg of the trityl were dissolved in 24.15\,mg water (deionized water from the lab) and 80.93\,mg D$_2$O (Sigma Aldrich, $\geq$ 99.85 atom \% D). This gives a $\sim$ 1.2\,$\mu$M solution of trityl dissolved in a D$_2$O:H$_2$O matrix of 3:1 by volume. From that resulting solution, 48.6\,$\mu$L were added to 99.92\,mg glycerol-d$_8$ (Sigma Aldrich, $\geq$ 98 atom \% D by Chromatography Purity) to give the desired 5\,mM solution of trityl in DNP juice. 40\,$\mu$L of that final solution were added to a 3\,mM OD quartz capillary. All masses were weighed in with a AT261 Delta Range scale with a precision of 0.01\,mg. A pipetman from GILSON (P100) with volumes adjustable to 0.1\,$\mu$L precision was used to transfer and measure the volumes.\newline
\textbf{5\,mM trityl in gly-d$_8$:D$_2$O (6:4)}\newline
The sample with a fully deuterated matrix was prepared inside a glove box. All chemicals and equipment were placed in the glove box 24 hours prior to sample preparation to prevent contamination from residual surface moisture.\newline
1.7\,mg of the trityl were dissolved in 112.5\,mg D$_2$O (Cambridge Isotope Laboratory, $\geq$ 99.85 atom \% D, sealed glass ampule). This gives a $\sim$ 1.2\,$\mu$M solution of trityl dissolved in a D$_2$O. From that resulting solution, 48.6\,$\mu$L were added to 100.2\,mg glycerol-d$_8$ (Sigma Aldrich, $\geq$ 98 atom \% D by CP, freshly opened in the glove box) to give the desired 5\,mM solution of trityl in a gly-d$_8$:D$_2$O (6:4 by volume) matrix. 40\,$\mu$L of that final solution were added to a 3\,mM OD quartz capillary inside the glove box. The quartz capillary was then transferred under inert condition and sealed using a Schlenk line. All masses were weighed in with a Mettler Toledo scale with a precision of 0.1\,mg. As GILSON PIPETMAN (P100 and P20) with volumes adjustable to 0.1\,$\mu$L precision were used to transfer and measure the volumes.
\newline
\textbf{5\,mM trityl in gly-d$_8$:H$_2$O (6:4)}\newline
1.68\,mg of the trityl were dissolved in 101.34\,mg H$_2$O (deionized water from the lab). This gives a $\sim$ 1.2\,$\mu$M solution of trityl dissolved in a H$_2$O. From that resulting solution, 48.6\,$\mu$L were added to 108.36\,mg glycerol-d$_8$ (Sigma Aldrich, $\geq$ 98 atom \% D by CP) to give the desired 5\,mM solution of trityl in a gly-d$_8$:H$_2$O (6:4 by volume) matrix. 40\,$\mu$L of that final solution were added to a 3\,mM OD quartz capillary. All masses were weighed in with a Mettler Toledo AX205 scale with a precision of 0.01\,mg. A GILSON PIPETMAN (P100) and GILSON (P20) with volumes adjustable to 0.1\,$\mu$L precision were used to transfer and measure the volumes.
\newline
\textbf{100\,$\mu$M trityl in gly-d$_8$:D$_2$O:H$_2$O (6:3:1)}\newline
0.33\,mg of the trityl were dissolved in 242.94\,mg water (deionized water from the lab) and 810.24\,mg D$_2$O (Sigma Aldrich, $\geq$ 99.85 atom \% D). This gives a $\sim$ 0.25\,$\mu$M solution of trityl dissolved in a D$_2$O:H$_2$O matrix of 3:1 by volume. From that resulting solution 486\,$\mu$L were added to 1000.2\,mg glycerol-d$_8$ (Sigma Aldrich, $\geq$ 98 atom \% D by CP) to give the desired 100\,$\mu$M solution of trityl in DNP juice. 40\,$\mu$L of that final solution were added to a 3\,mM OD quartz capillary. All masses were weighed in with a Mettler Toledo AX205 scale with a precision of 0.01\,mg. A GILSON PIPETMAN (P100) with volumes adjustable to 0.1\,$\mu$L precision was used to transfer and measure the volumes.\newline

All experiments were conducted at 80 K and an external magnetic field of around 0.35 T (X-band, $\nu_{\mathrm{0,e}}\sim$ 9.8 GHz, $\nu_{\mathrm{0,H}}\sim$ 14.83\,MHz). The sample was flash frozen in liquid nitrogen before placed in the resonator.

\subsection{EPR and NMR measurements}\label{subsec:EPR_NMR}
All experiments were acquired on a home-built X-band spectrometer similar to the spectrometer described in Ref. \cite{Doll2017}. A constant temperature of 80 K was achieved with a closed-cycle cryogen-free cryostat from Cryogenic Limited. An arbitrary waveform generator (Keysight model M8190A) was used to generate mw pulses and the pulses were amplified using a 1 kW traveling wave tube (TWT) amplifier. The Hahn echoes were recorded using a digitizer running at 1.8 GSa/s (SP devices ADQ412) for the measurements of the samples 5\,mM trityl in gly-d$_8$:D$_2$O:H$_2$O (6:3:1), 5\,mM trityl in gly-d$_8$:D$_2$O (6:4) and 5\,mM trityl in gly-d$_8$:H$_2$O (6:4). The sample 100\,$\mu$M trityl in gly-d$_8$:D$_2$O:H$_2$O (6:3:1) was measured using a digitizer running at 10 GSa/s (ADQ7DC). A Bruker EN4118X-MD4 resonator was used with a home-built external rf tuning and matching box. NMR pulses were generated with an OpenCore spectrometer. \cite{Takeda2007,Takeda2008} A schematic representation of a general DNP experiment used in this work can be seen in Fig. \ref{fig:RevDNPPulseSeq}. The general sequence starts with a NOVEL DNP contact consisting of a $\frac{\pi}{2}$-pulse of 6 ns length and a digital amplitude 1 (max. amplitude) and a spin lock of length $t_{\mathrm{SL}}$. In most of the experiments $t_{\mathrm{SL}}$ was set to 4000 ns. The optimal amplitude and length of the spin lock was determined by a single DNP contact followed by a Hahn echo detection as described in Section E.6. of the ESI\dag. After the first DNP contact the AWG was delayed for $t_{\mathrm{del}}$. During this delay pulses on the proton channel were applied using a BLAX300RS amplifier, where the maximum output power is 300 W corresponding to a relative amplitude of 100. A detailed description of the pulse scheme on the proton channel is given below. After the delay $t_{\mathrm{del}}$ the electron spin was saturated by a train consisting of five pulses of 10 ns length spaced by 1800 ns. This step is necessary to ensure that the proton spins are more strongly polarized than the electron spin and thus to facilitate the reverse DNP transfer (second spin lock). The reverse DNP was started 1800 ns after the last saturation pulse. The electron signal was then read out by a Hahn echo with $\tau$ = 600 ns and a $\pi$ of 12 ns length and a digital amplitude of 1. Each single-shot experiment was followed by a $^{1}$H saturation pulse train consisting of eleven $100^{\circ}$ pulses spaced by 1 ms to destroy any polarization left on the protons. The frequency of the saturation pulses was always on resonance with respect to tuning and matching. The shot repetition time was set to 15 ms $\sim5T_{\mathrm{1,e}}$. 
\newline
\textbf{Electron-detected proton spectra}\newline
To record the electron-detected proton spectra as described in Section \ref{subsec:HSpectrum} the nutation pulse (Gauss or rectangular pulse) on the proton channel was placed in the middle of the delay as shown in Fig. \ref{fig:RevDNPPulseSeq} A) for a variable-amplitude Gauss pulse. The carrier frequency of the nutation pulse $\nu_{\mathrm{rf,2}}$ was swept from 14.14\,MHz to 15.10\,MHz. For each frequency the rf coil in the resonator was tuned and matched accordingly. An overview of the minimum of the dip after tuning and matching is shown in Fig. S113 and S114 in the ESI\dag. In case of a rectangular nutation pulse the relative amplitude of the pulse was kept constant at 30 corresponding to 100 kHz and the length of the pulse was varied from 0 to 9\,$\mu$s in 0.5\,$\mu$s steps. In case of the Gauss pulse, the pulse length was kept constant at 10\,$\mu$s and the relative amplitude was varied from 0 to 40 in 2.5 steps. The minimum of the amplitude nutation trace was read out for $\nu_{\mathrm{rf,2}}=14.89$\,MHz corresponding to the center of the electron-detected proton spectra i.e. amplitude nutation trace with the largest inversion. For the variable-amplitude Gauss pulse the minimum was at a relative amplitude of 25 with and without electron decoupling. For the rectangular nutation pulse the minimum of the amplitude nutation trace was at $\tau_{\mathrm{nut}}=$ 3.5\,$\mu$s without electron decoupling and at $\tau_{\mathrm{nut}}=$ 4\,$\mu$s with electron decoupling applied during the nutation pulse. The electron-detected proton spectrum was then obtained by reading out the data points at the minimum position determined for $\nu_{\mathrm{rf,2}}=14.89$\,MHz. The spectra were baseline corrected to the signal corresponding to $\nu_{\mathrm{rf,2}}=14.14$\,MHz which did not show any nutation. They were afterwards multiplied by -1 to obtain a positive signal. Electron decoupling sequence consisted of 99 $\pi$ pulses (12 ns pulses at digital scale of 1) spaced by 80.2 ns and was applied during the nutation pulse (see Fig. S8 in the ESI\dag). Experiments were recorded for $t_{\mathrm{del}}=$ 15\,$\mu$s an 800\,$\mu$s.
\newline
\textbf{Detection of proton spin diffusion towards the bulk}\newline
To observe the proton spin diffusion into the bulk for different carrier $\nu_{\mathrm{rf,2}}$ the Gauss pulse was placed at the end of the delay (see blue Gauss pulse in Fig. \ref{fig:RevDNPPulseSeq} B)). The length of the Gauss pulse at the end was kept constant at 10\,$\mu$s and the relative amplitude of this pulse was swept from 0 to 40 in 2.5 steps. The delay $t_{\mathrm{del}}$ was varied from 15\,$\mu$s to 3000\,$\mu$s. The spin diffusion spectra were then obtained by the minimum of the amplitude nutation traces at a relative amplitude of 25 (equal to the electron detected proton spectra). The NMR coil of the resonator was tuned and impedance matched following the same procedure as outlined for the electron-detected proton spectra. 
\newline
\textbf{Electron spin diffusion imprinted on the proton spectrum}\newline
The pulse sequence containing two Gauss pulses on the proton channel is schematically shown in Fig. \ref{fig:RevDNPPulseSeq} C). For all experiments $t_{\mathrm{del}}=$ 800\,$\mu$s was fixed. The blue pulse was the detection pulse (Inversion pulse). The length of this pulse was set to 10\,$\mu$s and the amplitude was set to the minimum of the amplitude nutation trace recorded at $t_{\mathrm{del}}=$ 800\,$\mu$s in the experiment with only one Gauss pulse at the end. The NMR coil of the resonator was tuned and matched on resonance with respect to the carrier frequency $\nu_{\mathrm{rf,2}}$ of the detection pulse. The frequency $\nu_{\mathrm{rf,2}}$ was scanned through the proton spectrum from 14.14\,MHz to 15.10\,MHz. The red Gauss pulse served as an hole-burning pulse. The frequency $\nu_{\mathrm{rf,1}}$ was fixed to $14.69$, $14.89$ or $15.04$\,MHz and the pulse length was set to 10\,$\mu$s. The space $\Delta t$ between the two Gauss pulses was varied from 5\,$\mu$s to 750\,$\mu$s. The amplitude of red Gauss pulse $a_1$ was varied for every single $\Delta t$ to ensure optimal hole burning performance also for large frequency differences between the carrier frequencies of the two Gauss pulses i.e. $\nu_{\mathrm{rf,1}}$ and $\nu_{\mathrm{rf,2}}$. The minimum or maximum of the amplitude traces for a fixed $\Delta t$ were extracted and examples for all four different samples can be found in Sections C.1. - C.5. of the ESI\dag. The obtained spectra were zero-corrected to the data point at 14.14\,MHz, inverted and compared to the data with $a_1=0$, i.e. in the absence of an hole burning pulse.

The NMR spectra without any DNP involved were recorded with only pulses applied to the proton channel. The proton spectra were recorded by a solid echo consisting of two $90^{\circ}$ pulses of 2.5\,$\mu$s length separated by $t_{\mathrm{SE}} = 20\;\mu$s. For the solid echo a eight step phase cycle was used with $\{x,x,y,y,- x, - x, - y, - y\}$ for the first pulse and detection and $\{y,- y,x,- x, y, - y, x, - x\}$ for the second pulse. A dwell time of 4\,$\mu$s and a number of 1024 sampling points was used to sample the solid-echo signal. A saturation recovery pulse sequence with adjustable delay $\tau_{\mathrm{delay}}$ was used to record the $T_{\mathrm{1,H}}$ of the proton spin. The reference experiment was recorded with $\tau_{\mathrm{delay}} = 180\;\mathrm{s}\approx5\cdot T_{1,n}$. 
To process the NMR data, a cosine-squared apodization function was used and time-domain data was zero-filled to twice the number of recorded data points. After Fourier transformation of the resulting time-domain data, the peak was fitted by a Lorentzian-Gauss blend.

\printbibliography

\end{document}